\documentclass[letterpaper,twocolumn,10pt]{article}
\usepackage{usenix}

\usepackage{tikz}
\usepackage{amsmath}

\usepackage{filecontents}

\usepackage{comment}
\usepackage{threeparttable}
\usepackage{soul}
\usepackage{multirow}
\usepackage{graphicx}
\usepackage{caption} 
\begin{document}
%-------------------------------------------------------------------------------

%don't want date printed
\date{}

\title{\Large \bf A Broad Comparative Evaluation of Software Debloating Tools}

\author{
{\rm Michael D. Brown}\\
Trail of Bits
\and
{\rm Adam Meily}\\
Trail of Bits
\and
{\rm Brian Fairservice}\\
Grammatech, Inc.
\and
{\rm Akshay Sood}\\
Grammatech, Inc.
\and
{\rm Jonathan Dorn}\\
Grammatech, Inc.
\and
{\rm Eric Kilmer}\\
Trail of Bits
\and
{\rm Ronald Eytchison}\\
Trail of Bits
} % end author

\maketitle

%un-comment to turn off page numbers for submission
%\pagestyle{empty}

\begin{abstract}
  Software debloating tools seek to improve program security and performance by removing unnecessary code, called bloat. While many techniques have been proposed, several barriers to their adoption have emerged. Namely, debloating tools are highly specialized, making it difficult for adopters to find the right type of tool for their needs. This is further hindered by a lack of established metrics and comparative evaluations between tools. To close this information gap, we surveyed 10 years of debloating literature and several tools currently under commercial development to taxonomize knowledge about the debloating ecosystem. We then conducted a broad comparative evaluation of 10 debloating tools to determine their relative strengths and weaknesses. Our evaluation, conducted on a diverse set of 20 benchmark programs, measures tools across 12 performance, security, and correctness metrics.

  Our evaluation surfaces several concerning findings that contradict the prevailing narrative in the debloating literature. First, debloating tools lack the maturity required to be used on real-world software, evidenced by a slim 22\% overall success rate for creating passable debloated versions of medium- and high-complexity benchmarks. Second, debloating tools struggle to produce sound and robust programs. Using our novel differential fuzzing tool, DIFFER, we discovered that only 13\% of our debloating attempts produced a sound and robust debloated program. Finally, our results indicate that debloating tools typically do not improve the performance or security posture of debloated programs by a significant degree according to our evaluation metrics. We believe that our contributions in this paper will help potential adopters better understand the landscape of tools and will motivate future research and development of more capable debloating tools. To this end, we have made our benchmark set, data, and custom tools publicly available.

\end{abstract}

\section{Introduction}

Software debloating is an emerging research area focused on improving the security and performance of programs by removing unnecessary code  (i.e., \emph{bloat}) in the form of unnecessary features or extraneous library code. Bloat is pervasive in modern programs due to entrenched software engineering practices (e.g., design emphasis on reusable code modules) and tendencies (e.g., scope creep). To counter this, several debloating techniques targeting various stages of the software life cycle have been proposed. However, their practical adoption has not kept pace with research for several reasons. 

First, debloating tools are highly specialized. Each tool targets a different kind of bloat, type of software (compiled, interpreted, kernels, etc.), stage of the life cycle (source, IR, binary, etc.) and has a unique analysis method. This makes it challenging for adopters to select the right category of tools for their specific problem. Second, the literature lacks established metrics resulting in unclear, incomplete, and potentially misleading claims regarding how different methods improve performance and security. This occurs because authors borrow or create their own metrics, which has been shown to be problematic \cite{GSA, razor}. Ultimately, it is difficult for adopters to understand what benefits to expect as it is not immediately apparent which metrics are useful or relevant. This is exacerbated by superficial explorations of soundness risks in the literature, making it similarly difficult for adopters to understand debloating's trade-offs and consequences. Finally, the few comparative evaluations present in the literature are limited to a small number of tools, making it difficult for adopters to determine which tools are best in class. 

While broad studies on bloat prevalence \cite{multi-os, bloatfactors, deadmethods}, debloating trade-offs \cite{covaf}, and security metrics \cite{GSA, lessismore, sludge} have been published, the largest comparative evaluation of software debloating tools to date \cite{debloatbencha} is limited to only four tools, each of a separate type. This permits comparing individual tools across categories and how well they compose, however the quantity and diversity of tools is not sufficient to compare tools within categories or draw conclusions about the relative merits of different types of debloaters. Also, the evaluation uses only four metrics and may fail to detect important negative impacts of debloating such as excessive runtime or memory consumption in debloated programs.

\textbf{Motivation.} To date, there has been no holistic review, classification, analysis, and evaluation of the tools, metrics, benchmarks, and use cases in the software debloating ecosystem. We close this knowledge gap by conducting a survey and comparative evaluation of software debloating tools that seeks to answer the following motivating research questions:

\begin{enumerate}
    \itemsep=0em
    \item How can debloating tools be classified?
    \item Which security, soundness, and performance metrics are useful? What new metrics are needed? 
    \item Which benchmarks and configurations provide a holistic evaluation of debloating tools?
    \item How do tools within the same class compare?
    \item How do debloating tools compare across classes?
\end{enumerate}

\textbf{Summary of Contributions.} In this paper, we first present the findings from our survey of debloating literature in Section \ref{sec:survey}. We then describe our experimental methodology to evaluate debloating tools and their products in Section \ref{sec:methodology}. We present the results of our evaluation in Section \ref{sec:results}, and discuss the key findings of our study in Section \ref{sec:discussion}.

\section{Survey of Debloating Techniques}
\label{sec:survey}
To conduct our survey, we first searched Google Scholar for articles on software debloating, customization, partitioning, and specialization. Next, we triaged each search result and it's references to identify duplicates and ensure relevance. We included refereed works from all venues in our study, provided they made a meaningful contributions to debloating discourse. The most common reason we excluded a search result for lack of relevance was for making only a passing and generic reference to software debloating, for example by referring to it as a related research area. Ultimately, we selected over 70 scholarly articles dating back to 2013 for our survey. Additionally, we included various commercial and work-in-progress techniques at various stages of development to fully cover the field's breadth. A number of these works have not been made publicly available; the authors of these works provided us with the information, technical artifacts, and support necessary for our survey (and subsequent evaluation). 

As we used an intentionally broad definition of "software debloating", we included many works detailing debloating methods for technical artifacts not traditionally considered software such as containers \cite{cimplifier, debloatbenchc}, operating systems and their APIs \cite{confine, facechange, C2C, cozart}, test cases \cite{creduce, jreduce, perses}, firmware \cite{decaf, utrimmer}, and build dependencies \cite{depclean}, among others \cite{asandebloat, polydroid, cachetor, slimium, leansym}. In the following sections, we establish a debloating taxonomy that defines and classifies types of bloat, debloating workflows, types of debloating techniques, metrics, and benchmarks we observed in our survey. 

\subsection{Types of Bloat}
In general, debloating tools try to remove or neutralize two different types of unnecessary code, which we define as Type I and Type II bloat. Type I bloat is \emph{universally unnecessary} and can be removed without impacting the program's behavior for all intended end uses. The most common manifestation of Type I bloat is library code that is dynamically loaded into a processes' address space at runtime but will never be called. Although less common, some debloating tools target Type I bloat in the form of operating system (OS) code (e.g., system calls) that a program does not call to prevent their malicious re-use. Dead and unreachable code is also Type I bloat, but is not heavily targeted by debloaters because compilers and interpreters are highly effective at removing it.

By contrast, Type II bloat is \emph{end-use dependent}: code may or may not be Type II bloat depending on how its user(s) uses the program. Type II bloat typically takes the form of unnecessary program features, such as support for obscure or obsolete file formats in image processing software. Platform-dependent code that handles variations in OS and web browser interactions is another form of Type II bloat, however this kind of bloat is typically addressed by build-systems and scripting engines rather than debloating tools.

\subsection{Debloating Workflow and Use Cases}
\label{sec:workflow}
Despite myriad technical differences, virtually all debloating tools share a common high-level, six-stage user workflow:
\begin{enumerate}
    \itemsep=0em
    \item \textbf{Specification:} The user creates a tool-specific specification that outlines what program behaviors should be kept or eliminated.\footnote{Tools targeting only Type I bloat may not require specifications.}
    \item \textbf{Input:} The user provides the specification and program to the debloating tool.
    \item \textbf{Analysis:} The tool uses information in the specification to analyze the program and label code (e.g., as unnecessary, as part of a feature).
    \item \textbf{Transformation:} The tool modifies the program to separate bloat from useful code (e.g., cutting, reconstruction, attaching labels).
    \item \textbf{Output:} The tool produces a modified program that either does not contain bloat or contains the necessary labels to remove bloat at runtime.   
    \item \textbf{Validation:} The user validates that the modified program is sound and behaves as expected via manual testing or automated tools (e.g., fuzzer) of their choice. If the output is invalid, the user may restart the workflow.
\end{enumerate}

Despite a common high-level workflow, many tools define different use cases for potential end users. Tools targeting Type I bloat prescribe a fairly straightforward use case: eliminate as much bloat as possible. Tools targeting Type II bloat employ three common use cases. \textbf{Aggressive} use cases remove all functionality except for a single desired feature or software configuration. For example, aggressively debloating a file compression tool might be defined as removing all code not necessary to compress a file and write the result to disk. \textbf{Moderate} use cases retain several core and peripheral features and debloat the rest. A moderate debloating use case for a file compression tool would keep code for compressing, decompressing, and testing a compressed file's integrity. Fianlly, \textbf{Conservative} debloating \cite{GSA, carve} use cases are defined as those in which only a few peripheral features are removed.

\subsection{Classifying Techniques}
We sort the software debloating tools we surveyed into five distinct categories primarily differentiated by design choices made at various stages in the debloating workflow.

\subsubsection{Source-to-Source (S2S) Debloaters} 
S2S debloaters remove bloat directly from a program's source code. They primarily target Type II bloat, although they can be readily adapted to remove Type I bloat in libraries. Debloating information-rich source code has several advantages over other representations (i.e., IR or binary). First, developers have many options for the analysis stage; S2S debloaters can use code coverage \cite{chisel, covaf, debop, domgad}, fuzzing \cite{ancile}, build system analysis \cite{prat, proguard}, and manual annotation \cite{carve} techniques to produce feature-to-code mappings. Second, source code is easier to transform without compromising the program's soundness because the compiler has yet to strip away high-level information and replace it with machine semantics. The compiler is also useful in this case for detecting errors and optimizing away inefficiencies after source transformation. Third, debloated source code is still relatively straightforward to maintain, as relevant bug fixes or security patches can be applied before compilation. Despite these advantages, transforming source code can be technically complex due to the flexibility in its expression (e.g., syntactic sugar) and lack of compiler support for direct transformation of source code.

S2S debloaters generally require the user to have working technical knowledge about both the tool itself and the program they want to debloat. Creating specifications for these tools typically involves specifying test cases that exercise program behaviors that the user wants to retain. In practice, this requires a large number of test cases even for moderately complex programs due to the need for test cases that exercise edge cases, security checks, error handling code, etc. As a result, creating specifications for S2S debloaters can take on the order of several hours for non-trivial programs.

\subsubsection{Compiler-Based Specializers (CBS)}
CBS are similar to S2S debloaters in that they take source code as input and primarily target Type II bloat. Rather than debloat source directly, CBS \cite{trimmer, trimmer2, occam, occam2, deepoccam, lmcas, finegrainlibrary} use a compiler (e.g., LLVM) during the analysis and transformation stages to lower the source to IR, convert key input parameters or values to compile-time constants based on the user's debloating specification, and remove bloat via built-in compiler optimizations such as constant propagation, loop unrolling, and dead code elimination. This approach mitigates soundness concerns by using highly-reliable compiler passes instead of custom transformation routines. As a trade-off, CBS only support aggressive debloating use cases. While this means users generally do not need deep working knowledge of the program they are debloating, it makes them unsuitable for moderate or conservative use cases or scenarios where features are triggered by runtime inputs, such as network protocol commands. 

Because CBS are limited to aggressive use cases, they generally only require the user to know how the program will be executed on their system (i.e., a single command line invocation). As a result, creating debloating specifications for CBS typically takes less than an hour. However, there are some notable exceptions that may increase the amount of experience or time needed to configure and use the tool. Since the tools are tightly coupled with the compiler, in some cases the user may need to have working knowledge of the tools build configuration (i.e., compiler flags). Additionally, some tools may require the user to perform manual steps, such as identifying the boundary between input parsing and program logic if it cannot be automatically identified\cite{lmcas}.

\subsubsection{Binary-to-Binary (B2B) Debloaters}
B2B debloaters are conceptually similar to S2S debloaters, but differ in that they target program binaries (e.g., ELF, Java bytecode\cite{reddroid, jshrink, jred, xdebloat}). Binary-only approaches are error-prone and fraught with challenges, however they have the key advantage of being able to debloat legacy or closed source binaries. B2B debloaters have limited options for the analysis stage; they must rely on inherently imprecise binary analysis techniques such as execution tracing of test cases \cite{razor,binrec,binrec-repo, toss, hecate,binarycontrolflowtrimming, gtirb, binary-reduce, fblocker} and  binary lifting \cite{binrec,egalito,hecate,binarycontrolflowtrimming,gtirb,binary-reduce, fblocker} as well as heuristics \cite{razor, fblocker} for generating feature-to-code mappings. The transformation phase is similarly difficult due to the unique nature of binary formats (e.g., co-mingling of program instructions and data, indirect branching, etc.). Because fully recovering a binary is undecidable in the general case, B2B debloaters must carefully manipulate the input binary without violating their original layout by blanking out code with no-ops or invalid instructions \cite{toss, hecate} or adding a debloated version of the program as a new code section alongside the original code \cite{binrec, binrec-repo}. Ultimately, the key disadvantages of B2B approaches are also consequences of the limitations of binary analysis: they struggle to produce sound debloated binaries and require high quality binary recovery to be effective (e.g., Egalito\cite{egalito} requires position-independent code to ensure recovery).

B2B debloaters generally require working technical knowledge of the target program only. Like test case-based S2S debloaters, creating a specification requires a large number of test cases in practice to minimize soundness issues. For more complex programs, this may take users one or more hours to enumerate. Once processed, debloated binaries are difficult to maintain or further modify due to the limits of the transformation techniques mentioned previously. However, given that B2B debloaters are primarily used for legacy programs, this may not be a significant detriment to their use.

\subsubsection{Static Library (SL) Debloaters} 
SL debloaters focus on removing Type I bloat permanently from libraries consumed by a target program. They do not require a specification due to a straightforward analysis phase in which the tools statically compute the call graph for the target program and identify a precise set of required library functions. SL debloaters use a variety of transformation approaches including creating specialized libraries  via rewriting to remove \cite{BinTrimmer, lightweightbinarytailoring, jax, cov-based, jslim, mininode, minimalist, animatedead} or blank unused functions with no-ops or invalid instructions \cite{libfilter, nibbler}, fragmenting libraries \cite{bloatfactors}, rewriting executables to statically link library functions \cite{binary-reduce}, and replacing unnecessary library functions with stubs \cite{stubbifier}. By limiting their scope to Type I bloat, SL debloaters avoid many of the challenges experienced by other classes of tools. First, SL debloaters do not risk creating unsound programs, although programming methods such as reflection and indirection may not be handled by some tools. Second, SL debloaters require little or no human effort to use beyond specifying the target program. However, the consequence of this design choice is that these tools are not suitable for scenarios where Type II debloating is required. SL debloaters may also nullify operating system-wide efficiencies that rely on mapping a single copy of a shared library into multiple processes' address space.

\subsubsection{Runtime Debloaters} 
Runtime debloaters are similar to SL debloaters, the key difference being their transformation and output stages. To avoid permanent changes to libraries, runtime debloaters record required library functions as program metadata and interpose on the dynamic linking process during execution to debloat unnecessary library functions from the processes' memory. Approaches vary significantly, for example Piece-wise Compilation and Loading (PCL) \cite{pcl} embeds call graph information in binaries and uses a custom loader to rewrite unnecessary functions with invalid instructions at runtime; whereas BlankIt\cite{blankit} and Decker\cite{decker} take the opposite approach of loading only necessary functions into the program's runtime memory based on the current execution point. Other approaches use information from the build system such as configuration options\cite{config-driven} and package dependencies\cite{pacjam} to remove bloat. In addition to the advantages of SL debloaters, runtime debloaters also have the advantage of not disturbing code at rest, providing high confidence in the soundness of the result. However, this comes at significant price: high runtime overheads required to manage process memory.

\subsection{Analysis of Debloating Metrics}
\label{sec:analysis-metrics}
Software debloating literature does not have an established set of metrics for measuring success. Consequently there is significant variability in the metrics used across the works we surveyed. In most cases, metrics were adapted from similar program transformation techniques (e.g., code optimization), although in some notable cases new metrics have been introduced\cite{GSA, covaf, cbat, bloatfactors}. In total, we noted 30 different metrics which we sorted into three categories and describe in the remainder of this section.

\textbf{Performance} metrics measure the resource consumption of both the debloating tools themselves and the debloated programs they produce. In the latter case, performance is compared to that of programs that were not debloated to determine if debloating improves or worsens performance. Commonly used performance measures include: CPU runtime, memory overhead, static binary size, number of external resources required, and manual effort required to use tools. 

\textbf{Correctness and Robustness} metrics measure the stability and resistance to abuse of debloated programs. Typically, debloated programs are exercised by test suites or fuzzers to identify faulty outputs, crashes, and/or other undesirable outcomes. Interestingly, we observed that works using these metrics collected them for debloated programs only. Since correctness and robustness issues may be present in programs prior to debloating, such measures should be taken for both the original, non-debloated programs and their debloated variants to isolate issues introduced by a particular tool. Crystal and Casinghino's \cite{cbat} comparative binary analysis tool uses an SMT-based weakest precondition approach to demonstrate equivalence of two program binaries or highlight differences in their behavior, providing a high-assurance metric for correctness that can be applied to debloating scenarios.

\textbf{Security} metrics measure improvements to the program's security posture achieved through debloating. In our survey, we noted two main categories of security metrics: vulnerability elimination and code-reuse prevention.  Ostensibly, these metrics are used to claim desirable outcomes such as eliminating latent vulnerabilities, reducing the program's attack surface, and reducing the ease of code re-use attacks. However, the real-world utility of these metrics is subject to debate. 

Vulnerability elimination is typically evidenced in literature by showing that known vulnerabilities (e.g., those reported in MITRE's CVE database\cite{mitre-cve}) in benchmark programs are removed during debloating. While this metric demonstrates the \emph{possible} benefits of debloating, it is ultimately flawed because a debloating tool's ability to eliminate vulnerabilities in evaluation is a poor indicator of whether or not it can measurably eliminate vulnerabilities in practice.

First, suppose our goal is to eliminate known (i.e., n-day) vulnerabilities in a program or a code library. Debloating is not a realistic strategy because applying a patch is generally less costly, complex, invasive and risky than debloating. In the uncommon case that there is no patch or mitigation available, debloating can only remove the vulnerability if it resides in code that the user does not need because debloating tools by design avoid altering required code. If the vulnerability is indeed resident in unneeded code, then removing may not be necessary unless it can be triggered by an attacker.  

Now, consider the more realistic case where our goal is to eliminate unknown (i.e., 0-day) vulnerabilities via debloating. In this case, the benefits of debloating are highly circumstantial. Unknown vulnerabilities will only be removed if they happen to reside in code the debloater determines is unnecessary. Also, we have no way of knowing if we successfully removed a vulnerability or not. Thus it is a best effort strategy only. For both known and unknown vulnerabilities, a debloater's ability to remove vulnerabilities is primarily dependent upon what code the user needs and the location of the vulnerability, not the characteristics of the debloating tool itself.

Finally, the use of this metric can be misleading, inaccurate, or incomplete. For example, Qian et al. \cite{razor} showed that prior work claiming vulnerability elimination in Type II bloat \cite{chisel} also \emph{reintroduced} other historical vulnerabilities. Also, PCL \cite{pcl} reports eliminating known vulnerabilities in Type I bloat as a benefit of their tool. However, these vulnerabilities are in library functions that are statically unreachable from the program. As such, they are exploitable only under extreme or contrived scenarios. 

Attack surface reduction and code-reuse prevention metrics used in the literature largely focus on code-reuse gadgets as a unit of measure. Code reuse gadgets are small snippets of chain-able code present in the vulnerable program an attacker can use to implement an exploit in situations where code injection defenses prevent them from executing shell code directly. A common metric used to demonstrate security improvement is a reduction in the total number of gadgets available to the attacker. However, Brown and Pande \cite{GSA} have shown this to be unsuitable due to the propensity of many debloaters to introduce new gadgets at a high rate despite reducing total gadget count and the likely possibility that large reductions in gadget counts fail to impose barriers on attackers creating exploit chains. In turn, Brown and Pande propose qualitative gadget set metrics for measuring costs imposed on attackers exploit (e.g., gadget set expressivity, composability, special capabilities, and locality) and a static analysis tool for computing them, GSA (Gadget Set Analyzer).

\subsection{Benchmark Programs}

The works we surveyed used a variety of different benchmarks to evaluate their tools. The majority of works employed benchmark sets that are commonly used in program analysis literature such as GNU Coreutils\cite{coreutils}, SPEC CPU 2006/2017 \cite{spec2006, spec2017} and DaCapo \cite{dacapo}. Additionally, several works created and subsequently made available their own benchmark sets such as CHISELBench \cite{chiselbench} and OCCAM benchmarks \cite{occambench}, however it is worth noting that these sets contain many benchmark programs from Coreutils and SPEC. Generally, the majority of benchmark programs used are of low complexity: they have command line user interfaces, make limited use of multi-threading, network interfaces, etc., and often run to termination with only a single set of inputs. This is unsurprising, as these types of benchmark programs are amenable to the complex tracing and transformation operations required for debloating. Many works also evaluated their tools on moderate-complexity benchmarks such as \texttt{bfptd}, \texttt{curl}, and \texttt{httpd}, although with less volume and frequency. Such benchmarks are characterized by use of complex inputs, multi-threading, network sockets, etc. Although rare, some works\cite{razor, slimium} evaluated their tools on high-complexity software such as web browsers and document readers.

\section{Evaluation Methodology}
\label{sec:methodology}
In the following subsections, we detail our methodology for selecting the tools, configurations, metrics, benchmarks that we used in our subsequent evaluation.

\subsection{Tool Selection}

Due to the size and diversity of the software debloating ecosystem, we first scoped our evaluation to tools that support userland C/C++ programs and libraries for x86/x86-64 machines. This benchmark profile is the most widely supported and had the largest pool of candidate tools. In total, we identified 31 candidate tools after filtering predecessor tools (e.g., including OCCAMv2\cite{occam2} versus OCCAMv1 \cite{occam}). We obtained source code via a public repository or request to the authors for 24 tools, and were able to successfully build and run 17 of them. We expended significant effort to resolve issues for the seven failing builds, however we were not successful due to irreconcilable technical issues \cite{pcl,ancile,domgad,pacjam} or unresponsive authors\cite{finegrainlibrary,hecate,toss}. Of the remaining 17 tools, we further excluded one because it requires a commercial license for IDA Pro \cite{bloatfactors} and six others because they were not readily adaptable to new targets. The underlying causes for exclusion included one tool that was hardcoded to its benchmarks\cite{prat}, several tools that required time consuming and manual pre-processing steps for new benchmarks \cite{carve, blankit, decker, covaf}, and one tool with technical limitations \cite{binarycontrolflowtrimming}. 

Table \ref{tab:tools} lists the ten tools that were suitable for our evaluation. For each tool, we prepared an isolated environment (i.e., virtual machine or docker container) configured with the tool's latest supported OS and all necessary resources required to use the tool and manipulate our benchmark programs. Of note, three tools are specialized versions of academic tools developed by commercial firms. To avoid confusion, we hyphenate the commercial firm's abbreviation to the tool title: CHISEL-GT (GrammaTech, Inc.), BinRec-ToB (Trail of Bits), and LMCAS-SIFT (Smart Information Flow Technologies). We disclose that development of these commercial tools and the evaluation presented in this paper was directed and funded by the Office of Naval Research (ONR). Further, the the authors of this paper are also the creators of BinRec-ToB and CHISEL-GT. The creators of LMCAS-SIFT also provided their tool and support for this evaluation.

% \usepackage{multirow}
% \usepackage{graphicx}
\begin{table*}[t]
\rowcolors{2}{lightgray!50}{white}
\centering
\caption{Debloaters Selected for Evaluation}
\setlength{\tabcolsep}{6pt}
\label{tab:tools}
\resizebox{.98\textwidth}{!}{%
\begin{tabular}{|ccc|cccc|c|c|}
\hline
\textbf{Year} & \textbf{Cite} & \textbf{Name} & \textbf{Category} & \textbf{Bloat Type} & \textbf{Granularity} & \textbf{Spec. Type} & \textbf{Open Source}  \\ \hline 
2018 & \cite{chisel}  & CHISEL & S2S &  Type II  & Line & Test Cases          &                                 $\bullet$    \\
2023 & \cite{chisel}    & CHISEL-GT   & S2S & Type II & Line &  Test Cases         &                                  \\
2019 & \cite{razor} & RAZOR & B2B   &  Type II  & Basic Block & Test Cases     &                                 $\bullet$ \\
2022 & \cite{binrec, binrec-repo}  & BinRec-ToB  &  B2B  & Type II  & Basic Block &  Test Cases         &                    $\bullet$         \\
2023 & \cite{binary-reduce}       & GTIRB Binary Reduce (Dynamic)  & B2B &  Type II &  Basic Block  & Test Cases &                                        \\ \hline
2021 & \cite{trimmer, trimmer2}        & Trimmer (v2) & CBS & Type II  & Instruction & Command           &                         $\bullet$       \\
2022 & \cite{occam, occam2}       & OCCAM (v2)  & CBS & Type II  & Instruction & Command   &                      $\bullet$     \\
2023 & \cite{lmcas}        & LMCAS-SIFT   & CBS &  Type I/II & Instruction & Command  &                                        \\ \hline
2023 & \cite{binary-reduce}     &  GTIRB Binary Reduce (Static)  & SL  & Type I  & Function   & None          &                            \\
2019 & \cite{libfilter}    & Libfilter  &  SL & Type I & Function & None     &               $\bullet$   \\ \hline
        
\end{tabular}%
}
\vspace{-.5em}
\end{table*}

\subsection{Benchmark Selection}
\label{sec:benchmarks}

We selected 20 C/C++ benchmark programs (Table \ref{tab:benchmarks}) that vary in size, complexity, and functionality. To represent low-complexity benchmarks, we use CHISELBench \cite{chiselbench} as it is commonly used in debloating literature and provides a common point of comparison for prior work. We further add six medium-complexity and four high-complexity benchmarks sourced from other benchmark sets \cite{occambench, rewriter-eval}. To build reference 64-bit ELF binaries for these benchmarks, we used Clang/LLVM v10 on Ubuntu Linux v20 with two build options specified: position-independent code (\texttt{-fPIC}) and optimization level 3 (\texttt{-O3}). However, due to various tool- and benchmark-specific limitations, we used different methods to build some reference binaries. We built reference binaries for use with RAZOR and CHISEL on Debian Buster. Building \texttt{ImageMagick} required using GCC v9.4 (Ubuntu) and GCC v8.3 (Debian Buster) instead of Clang/LLVM. Finally, we built the reference binaries for BinRec-ToB as 32-bit ELF binaries as it does not support 64-bit programs. 

% \usepackage{multirow}
% \usepackage{graphicx}
\begin{table*}[t]
\rowcolors{2}{lightgray!50}{white}
\centering
\caption{Evaluation Benchmarks}
\setlength{\tabcolsep}{6pt}
\label{tab:benchmarks}
\begin{threeparttable}
\begin{tabular}{|c|c|c|c||c|c||c|c|}
\hline

\multicolumn{4}{|c||}{\textbf{Low Complexity (CHISELBench\cite{chiselbench})}} & \multicolumn{2}{c||}{\textbf{Medium Complexity}} & \multicolumn{2}{c|}{\textbf{High Complexity}} \\ \hline
\textbf{Benchmark} & \textbf{Size} (KLOC) & \textbf{Benchmark} & \textbf{Size} & \textbf{Benchmark} & \textbf{Size} & \textbf{Benchmark} & \textbf{Size} \\ \hline

\texttt{bzip2 v1.0.5} & 12.2 &  \texttt{rm v8.4}  & 7.4 & \texttt{bftpd v6.1} & 4.7 &  \texttt{nmap v7.93} & 233.4 \\
\texttt{chown v8.2} & 7.3 &    \texttt{sort v8.16} & 14.7 & \texttt{wget v1.20.3} & 14.2 & \texttt{nginx v1.23.3} & 170.6  \\
\texttt{date v8.21} & 9.9 &   \texttt{tar v1.14} & 31.3 & \texttt{make v4.2} & 24.6 &    \texttt{pdftohtml v0.60} & 16.1 \\
\texttt{grep v2.19} & 23.8 &   \texttt{uniq v8.16} & 7.4 & \texttt{objdump v2.40} & 59.7 &   \texttt{ImageMagick v7.0.1} & 361.9 \\ 
\texttt{gzip v1.2.4} & 8.9 & & & \texttt{memcached v1.6.18} & 30.5 & & \\
\texttt{mkdir v5.2.1} & 5.1 & & & \texttt{lighttpd v1.4} & 89.7 & & \\ \hline
      
\end{tabular}%
\begin{tablenotes}
    %\scriptsize
      \item NOTE: Due to incompatibilities with LMCAS-SIFT, we were forced to use different minor versions of some benchmarks when building this tool's reference binaries: \texttt{coreutilsv8.26 (chown, date, mkdir, rm, sort, uniq)} and \texttt{gzip v1.12}.
    \end{tablenotes}
\end{threeparttable}

\vspace{-.5em}
\end{table*}

\subsection{Debloating Specifications}
\label{sec:debloat-specs}

Due to varying support for debloating use cases across tools, we used three strategies for creating debloating specifications. First, specifications are not required for SL debloaters as they target Type I bloat only. Second, we adopted a moderate (Section \ref{sec:workflow}) debloating use case for the S2S and B2B debloaters as this use case is most prevalent in the literature. Third, we adopted an aggressive use case for the CBS debloaters as this is the only use case they support. 

For each benchmark, we created a general, tool-agnostic debloating specification consisting of several core and peripheral features to retain (i.e., the moderate use case). We defined each feature with a descriptive name and one or more sample commands (i.e., test cases) for the benchmark that exercises the feature. We further winnowed these specifications to a single command to create an aggressive version of the moderate use case. Using the general specifications as a guide, we then created tool-specific configuration files for each benchmark program. In total, we created 160 distinct debloating specifications for use in our evaluation \footnote{ General debloating specifications and tool-specific configurations are available in our data repository (Section \ref{sec:availability}).}.

\subsection{Metric Selection}
We selected 12 metrics (Table \ref{tab:metrics}) for evaluating both the debloaters themselves and the debloated programs/libraries they produce. These metrics were drawn from those frequently used in the three categories outlined in Section \ref{sec:analysis-metrics}. Generally, we selected metrics that are computed from analysis of program binaries, as this is the only common program representation to all debloating tools (e.g., debloated code output by S2S or CBS debloaters can be compiled and measured against B2B debloated binaries). Of note, we chose not to use CVE removal as a security metric in our evaluation due to its poor predictive power. We provide details for how each metric is computed and our evaluation results in Section \ref{sec:results}.

% \usepackage{multirow}
% \usepackage{graphicx}
\begin{table*}[t]
\rowcolors{2}{lightgray!50}{white}
\centering
\caption{Metrics Selected for Evaluation}
\setlength{\tabcolsep}{6pt}
\label{tab:metrics}
\resizebox{.98\textwidth}{!}{%
\begin{tabular}{|c|c||c|c|}
\hline

\textbf{Name} & \textbf{Category} & \textbf{Name} &\textbf{Category} \\ \hline 
Run time (tool)  & Performance  &  Gadget Set Expressivity &  Security\\
Peak Memory Use (tool)  & Performance   &  Gadget Set Quality & Security    \\
Run time  & Performance  & Gadget Set Locality  & Security \\
Peak Memory Use & Performance   & Special Purpose Gadget Types Available & Security    \\
Static Binary Size  & Performance  & Executes Retained Functions  & Correctness / Robustness  \\
Number of Libraries Linked &  Performance   &   Errors / Crashes during Differential Testing &  Correctness / Robustness  \\
\hline

% Time Required to Generate Debloating Spec & Usability
% Time Required Harness Programs & Usability
% Operator Expertise Required & Usability
% Maintenance Difficulty &  Usability
      
\end{tabular}%
}
\vspace{-.5em}
\end{table*}

\section{Evaluation Results}
\label{sec:results}

We present the results of our evaluation organized by metric categories. First, we present performance results for the tools themselves, followed by performance, correctness, and security metrics calculated by analyzing the debloated artifacts they produce. Note that we discuss important findings and results for each metric in this section but leave a holistic and cross-category analysis for Section \ref{sec:discussion}.

\subsection{Tool Performance}

Our evaluation of debloating tool performance is limited to its \textbf{CPU runtime} and \textbf{peak memory usage} during execution. We exclude various other aspects of tool use (e.g., preparing the tool, harnessing the target program, deploying the debloated program, etc.) from our performance evaluation as they are highly dependent upon the user's technical expertise and familiarity with tool, target program, and deployment environment. To conduct our performance evaluation, we placed each tool's isolated execution environment, their relevant debloating specifications, and our benchmark programs (source code and reference binaries) on our dedicated testing virtual machine (VM). We configured the VM as a server-class machine with access to 24 3.2 GHz processors and 128 GB of memory.

Some manual effort was required for three tools before we could execute our tool performance tests. First, we discovered that CHISEL and CHISEL-GT were unable to run within a reasonable time frame of 48 hours for medium and high-complexity benchmarks as designed. CHISEL/CHISEL-GT requires many iterations to progressively identify unnecessary code in programs, each of which requires deleting some source code, compiling the resulting code, and executing the numerous test cases. As such, CHISEL and CHISEL-GT's debloating runtime is proportional to the runtime of the test cases that make up its specification. To evaluate these tools beyond the CHISELBench (i.e., low-complexity benchmarks), we worked with the authors of these tools to implement a parallelized test framework to improve its performance enough to meet our standard for reasonable run time (48 hours). In all, this engineering effort required 26 engineer hours to implement. 

Second, LMCAS-SIFT introduces the concept of a program "neck", which is defined as the boundary between a program's configuration logic (i.e., command line and/or config file parsing code) and main logic (i.e., the rest of the program). LMCAS-SIFT requires that a suitable neck be identified prior to debloating, and provides an automated placement tool for this purpose called the neck miner. The neck miner does not always identify the neck properly, leading to debloating failures. If the neck miner fails, the user must manually identify it. In our evaluation, the neck miner failed to automatically place the neck for \texttt{gzip}, \texttt{mkdir}, \texttt{sort}, \texttt{tar}, and \texttt{uniq}, and required us to manually place them. 

With our testing environment configured and manual harnessing complete, we then ran each tool ten times\footnote{Due to prohibitively long run times on the order of several CPU hours/days (even with customization), we ran several tools on certain benchmarks only once: 1) all CHISEL and CHISEL-GT performance tests, 2) OCCAM and TRIMMER on \texttt{objdump.}} on each compatible benchmark program recording the total CPU runtime and peak memory usage per run. We summarize the average run time in CPU minutes and peak memory use across all benchmarks at each level of complexity in Table \ref{tab:tools_perf}. Note that we include debloating operations that fail in these calculations. We exclude only benchmarks that are not compatible with the tool for reasons such as the tool does not support C++ code, multi-threaded programs, etc. The first set of columns in Table \ref{tab:tools_perf} indicate the number of compatible benchmarks per tool at each level of complexity, out of the total 10 low, 6 medium, and 4 high complexity benchmarks.

% Tables merged, don't use singles
%\input{Tables/tool_runtime}
%\input{Tables/tool_memory}
% \usepackage{multirow}
% \usepackage{graphicx}
\begin{table*}[t]
\rowcolors{4}{lightgray!50}{white}
\centering
\caption{Average Run Time (CPU minutes) and Peak Memory (MB) per Debloater}
\setlength{\tabcolsep}{6pt}
\label{tab:tools_perf}
%\resizebox{.98\textwidth}{!}{%
\begin{threeparttable}
\begin{tabular}{|c|c|c|c||c|c|c||c|c|c|}
\cline{2-10}
\multicolumn{1}{c|}{} & \multicolumn{3}{c||}{\textbf{\# Benchmarks}} & \multicolumn{3}{c||}{\textbf{Run Time}} & \multicolumn{3}{c|}{\textbf{Peak Memory}} \\ \hline
\textbf{Debloater} & \textbf{Low} & \textbf{Medium} & \textbf{High} & \textbf{Low} & \textbf{Medium} & \textbf{High} & \textbf{Low} & \textbf{Medium} & \textbf{High} \\ \hline
CHISEL & 9 & 6 & 4 & 282 & 224$^1$  & 324 & 306 & 91 & 181 \\
CHISEL-GT  & 9 & 4 & 3 & 4078 & 4058 & 13350 & 2145 & 1675 & 3011 \\
RAZOR & 10 & 5 & 4 & <1  & <1 & 10 & 32  & 31 & 71  \\
BinRec-ToB  & 9 & 3 & 3 & 5 & 7 &  8  & 1240 & 1251 &  1239 \\
Binary Reduce (Dyn.) & 10 & 6 & 4 & <1 & 2 &  3 & 173 & 853 & 1085 \\ \hline
Trimmer (v2) [Agg.] & 10 & 6 & 4 & 2 & 90 $^2$  &  18 & 3442 & 6518 & 5270  \\
OCCAM (v2) [Agg.]  & 10 & 6 & 4 & <1 & 150 $^2$  &  <1 & 73 & 5565  & 147 \\
LMCAS-SIFT [Agg.] & 10 & 5 & 1 & <1 & <1 & 6 & 458 & 625 & 2001 \\ \hline
Binary Reduce (Static) & 10 & 6 & 4 & 1 & 7 & 25 & 415 & 1577 & 3323  \\
Libfilter & 10 & 6 & 4 & <1 & <1 & <1 & 304 & 503 & 465 \\ \hline

\end{tabular}%
\begin{tablenotes}
      %\scriptsize
      \item $^1$ 15 excluding outlier \texttt{make} with 21 CPU hour runtime.
      \item $^2$ <1 excluding outlier \texttt{objdump} with 8.9 (Trimmer) and 15 (OCCAM) CPU hour runtimes.
      %\item $^3$ <1 excluding outlier \texttt{objdump} with 15 CPU hour runtime.
    \end{tablenotes}
\end{threeparttable}
%}
\vspace{-.5em}
\end{table*}

Excluding a notable outlier (\texttt{objdump}), all debloaters except CHISEL and CHISEL-GT took 25 minutes or less on average to complete. CHISEL and CHISEL-GT took on the order of CPU hours and CPU days to run, respectively. This is due to their rather inefficient and indirect method of mapping features to source code by iterative deletion and compilation/testing. With respect to memory consumption, debloaters generally required memory comparable to the size of the target program. Across all tools, we observed a maximum average peak demand of 6.4GB of memory. Finally, we observed very little variability in tool CPU runtime and peak memory usage across all benchmarks and trials. In all cases, the standard deviation across each tool's ten performance trials on a given benchmark was 1-2 orders of magnitude below the average value.

\subsection{Program Performance}
\label{sec:bench_perf}
Software debloating tools seek to improve, or at the very least avoid negatively impacting a program's performance when removing its unnecessary code. The primary expectation is that the debloated program will occupy less space both on disk and in run time memory, however it is not unrealistic to expect that debloated programs may also execute faster if they are structured in such a way where unnecessary code is frequently executed. Conversely, it is also possible that the mechanics of debloating operations may degrade run time. We present in this section four performance metrics collected on the debloated binaries and libraries that capture the performance impacts of using these tools.

\textbf{Static Binary Size.} We recorded the size of each successfully debloated binary on disk and compare it to the size of the original program's binary. Our calculations are modified for LMCAS-SIFT and the SL debloaters due to differences in their implementation. Libfilter produces new, slimmed copies of the benchmark's linked libraries; thus we compare the aggregate sizes of the debloated libraries against the originals. The debloated programs created by LMCAS-SIFT and Binary Reduce (Static) are statically linked binaries. For these we compare the statically linked binary's size to the aggregate size of the original binary and its dynamically linked libraries. 

We calculated size changes as a percentage, with values under 100\% indicating size reductions. We then averaged the changes in binary size across the three complexity levels, shown in Table \ref{tab:binary_size}. The first set of columns indicates the number of successfully debloated benchmarks for each tool at each complexity level. BinRec-ToB was not able to successfully debloat any of the benchmarks, and as such we exclude it from this and all future tables in our evaluation. 

% \usepackage{multirow}
% \usepackage{graphicx}
\begin{table*}[t]
\rowcolors{4}{lightgray!50}{white}
\centering
\caption{Average Change in Static Binary Size}
\setlength{\tabcolsep}{6pt}
\label{tab:binary_size}
%\resizebox{.98\textwidth}{!}{%
\begin{threeparttable}
\begin{tabular}{|c|c|c|c||c|c|c|}
\cline{2-7}
 \multicolumn{1}{c|}{} & \multicolumn{3}{c||}{\textbf{\# Benchmarks}} & \multicolumn{3}{c|}{\textbf{Static Binary Size}} \\ \hline
\textbf{Debloater} & \textbf{Low} & \textbf{Medium} & \textbf{High} & \textbf{Low} & \textbf{Medium} & \textbf{High} \\ \hline 
CHISEL & 9 & 6 & 2 & 31.6\% & 98.4\%  & 100\% \\
CHISEL-GT & 8 & 3 & 2 & 80.9\% & 79.7\% & 96.6\% \\
RAZOR & 10 & 3 & 3 & 117.2\% & 101.5\% & 107.3\%  \\
Binary Reduce (Dynamic) & 10 & 5 & 4 & 54.4\% & 38.6\% & 76.4\% \\ \hline
Trimmer (v2) [Agg.] & 7 & 4 & 4 & 68.1\% & 60.6\% & 79.6\%  \\
OCCAM (v2) [Agg.]  & 10 & 6 & 4 & 80.5\% & 74.7\%  & 93.9\% \\
LMCAS-SIFT [Agg.]$^1$  & 7 & 2 & 1 & 6.4\% & 8.3\% & 33.0\% \\ \hline
Binary Reduce (Static)$^1$ & 10 & 5 & 1 & 22.2\%  & 46.5\%  & 49.0\% \\
Libfilter & 10 & 4 & 1 & 101.3\% & 101.1\%  & 101.5\% \\ \hline

\end{tabular}%
\begin{tablenotes}
    %\scriptsize
      \item $^1$ Size calculated over executable and libraries because output is a statically linked binary.
    \end{tablenotes}
\end{threeparttable}
%}
\vspace{-.5em}
\end{table*}

Our evaluation reveals several interesting results. RAZOR and Libfilter produced debloated binaries that are larger on average across all complexity levels. This is not surprising, as RAZOR stitches the debloated version of the program into a new code section of the original binary and marks the original code section as non-executable rather than actually removing it. Similarly, Libfilter rewrites unreachable functions as \texttt{HLT} (halt) instructions rather than removing them. We note that reducing binary size is not a stated goal of either tool, and this metric was not evaluated by their authors. Further, the transformation methods used by these tools can be replaced with other methods that reduce binary size rather than increase it without fundamental changes to their approaches.

Generally, other tools achieved large size reductions for low complexity benchmarks but were noticeably less effective on high complexity benchmarks. This, combined with poor success rates on high-complexity benchmarks (only 3 of 10 tools successfully processed all four) suggests that these tools are likely over-fit to low- and medium-complexity benchmarks.  

\textbf{Number of Linked Libraries}
In addition to recording reduction in binary size, we also recorded the number of external libraries linked by our benchmarks before and after debloating to determine if removing Type II bloat in the main executable eliminated the need for one or more libraries. We exclude LMCAS-SIFT and the SL debloaters from this evaluation since they operate on libraries directly. Our results are intended to be anecdotal only because eliminating libraries via debloating is largely dependent upon the specification.

Binary Reduce (Dynamic), CHISEL and CHISEL-GT were generally successful in eliminating one or more libraries for medium- and high-complexity benchmarks. Such observations are expected based on their design, however we observed something unexpected: TRIMMER and OCCAM \emph{introduced} new libraries into debloated programs. The introduced libraries are typically \texttt{libc++}, \texttt{libgcc}, and \texttt{libm}, and are likely the result of changes the tool makes to the program with its custom LLVM compiler transformation passes. These libraries are quite large and computationally complex. Depending on the user's goals for debloating (e.g., minimizing code size or re-usability), introducing new dependencies like this may prove counterproductive.

\textbf{Program Run Time and Peak Memory Use.}
To determine how debloating processes impact the performance of debloated programs, we created test suites for each benchmark that exercise retained functionality according to our specifications (moderate and aggressive). These test suites are designed to be strenuous and long-running where possible. We ran each benchmark's reference binaries and each successfully debloated binary against these test suites ten times, averaging the total run time in CPU seconds and peak memory consumption in MB across the trials. In many cases, the debloated programs could not successfully complete the performance test due to crashes, infinite execution, or run-time errors. We excluded these failures from our calculations. 

We calculated changes in run time and peak memory as a percentage, with values under 100\% indicating reductions. We then averaged the changes across the three complexity levels, shown in Table \ref{tab:bench_perf}. The first set of columns indicates the number of debloated benchmarks that successfully completed the performance tests for each tool at each complexity level. Note that we exclude 3 medium and 2 high complexity benchmarks for CHISEL from these counts and our calculations because CHISEL "successfully" processed these benchmarks by making no modifications to the source code.

% Tables merged, don't use singles
%\input{Tables/bench_runtime}
%\input{Tables/bench_memory}
% \usepackage{multirow}
% \usepackage{graphicx}
\begin{table*}[t]
\rowcolors{4}{lightgray!50}{white}
\centering
\caption{Average Change in Benchmark Run Time and Peak Memory Performance}
\setlength{\tabcolsep}{6pt}
\label{tab:bench_perf}
%\resizebox{.98\textwidth}{!}{%
\begin{threeparttable}
\begin{tabular}{|c|c|c|c||c|c|c||c|c|c|}
\cline{2-10}
\multicolumn{1}{c|}{} & \multicolumn{3}{c||}{\textbf{\# Benchmarks}} & \multicolumn{3}{c||}{\textbf{Run Time}} & \multicolumn{3}{c|}{\textbf{Peak Memory}} \\ \hline
\textbf{Debloater} & \textbf{Low} & \textbf{Medium} & \textbf{High} & \textbf{Low} & \textbf{Medium} & \textbf{High} & \textbf{Low} & \textbf{Medium} & \textbf{High} \\ \hline 
CHISEL & 7 & 2 $^2$ & 0 $^2$ & 207\% $^1$ & 97.9\% & N/A & 96.7\% & 99.6\% & N/A \\
CHISEL-GT & 7 & 2 & 1 & 97.6\% & 100.7\% & 101.8\% & 99.6\% & 99.7\% & 100.4\% \\
RAZOR & 10 & 3 & 2 & 95.8\% & 106.3\% & 100.9\% & 99.8\% & \% 99.4 & 100.6\% \\
Binary Reduce (Dynamic) & 0 & 0 & 1 & N/A & N/A & 100.9\% & N/A & N/A & 99.9\%  \\ \hline
Trimmer (v2) [Agg.] & 6 & 1 & 2 & 135.8\% & 175.9\% & 138.6\% & 105.1\% & 106.9\% & 102.5\% \\
OCCAM (v2) [Agg.]  & 10 & 1 & 0 & 128.2\% & 175.7\% & N/A & 110.3\% & 101.4\%  & N/A\\
LMCAS-SIFT [Agg.]  & 7 & 0 & 0 & 103.4\% & N/A & N/A & 99.1\% & N/A & N/A  \\ \hline
Binary Reduce (Static) & 9 & 4 & 0 & 95.2\% & 101.0\% & N/A & 97.2\% & 98.2\% & N/A \\
Libfilter & 8 & 2 & 0 & 102.4\% & 100.4\% & N/A & 100.3\% & 100\% & N/A\\ \hline

\end{tabular}%
\begin{tablenotes}
      %\scriptsize
      \item $^1$ 95.4\% excluding outlier: 844.4\% performance on \texttt{gzip}
      \item $^2$ Excludes 3 medium and 2 high complexity benchmarks that were "successfully" processed, but left unmodified by CHISEL.
    \end{tablenotes}
\end{threeparttable}
%}
\vspace{-.5em}
\end{table*}

Our performance evaluation surfaces an important finding: we observed a high failure rate for medium and high-complexity benchmarks across all debloaters. Collectively, only 22\% of our attempts to debloat these benchmarks were successful, calling into question the utility of software debloating tools for complex programs. Further, the success rate on low-complexity benchmarks was also lower than expected. Despite CHISELBench being used extensively in the literature, only two tools, OCCAM and RAZOR successfully debloated all low-complexity benchmarks. Conversely, BinRec-ToB and Binary Reduce (Dynamic) failed to debloat any low complexity benchmarks. 

With respect to our performance metrics, we observed only a few instances where debloated programs performed noticeably worse than their unmodified counterparts. Excluding an outlier for CHISEL, only TRIMMER and OCCAM demonstrated consistent and significant (>4\%) negative impacts to run time or memory consumption across multiple complexity levels. The most significant of which are large (28\% to 76\%) increases in run time. This is somewhat unsurprising as these debloaters significantly alter the build chain and overall phase ordering of compiler optimizations to facilitate debloating. Phase ordering has significant effects on binary performance and is an open area of research\cite{kulkarni2009practical, kulkarni2012mitigating}.

\subsection{Program Security Posture}
Aside from performance improvements, debloating literature often cites improving a program's security posture (i.e., reducing its attack surface) as a chief motivating goal. However, concrete definitions of attack surface reduction vary widely and there is a lack of sensible and effective metrics for measuring it. Given the shortcomings of other commonly used measures (i.e., elimination of known vulnerabilities), we adopt four metrics that focus on code re-usability for our evaluation. We use GSA\cite{GSA} to analyze our debloated binaries and their unmodified counterparts and compute these metrics, which capture how readily an attacker can re-use the code present in the program and its linked libraries to manufacture an exploit in the presence of common security controls such as W$\oplus$X and ASLR. Without a priori knowledge of the actual vulnerabilities present in a program, these metrics provide a qualitative assessment of the hygienic security benefits of debloating.

\textbf{Gadget Set Expressivity and Quality}. The first two code re-usability metrics we evaluated are concerned with the aggregate properties of the functional code reuse gadgets available in a program binary. Attackers chain together functional code reuse gadgets like instructions in a programming language to write and launch an exploit without injecting code. Gadget set expressivity measures the collective expressive power of functional gadgets in a binary to determine if they are sufficient to launch practical exploits. GSA defines 11 different classes of functionality that must be met by gadgets in the set to meet this bar and reports expressivity as the number of satisfied classes. Gadget set quality measures the average chain-ability of gadgets in the set to determine how easily they can be composed. Gadgets useful to an attacker often contain machine instructions that are irrelevant for their purposes but may impose side constraints on creating a gadget chain (i.e., altering the stack pointer). GSA measures the quality of each gadget by using a starting score of 0.0 and adding to this score for each side constraint detected. Overall set quality is computed as the average score across all gadgets.

For both metrics, GSA compares set expressivity and quality in the debloated binary versus the original and reports the change in value. For LMCAS-SIFT and the SL debloaters, we include libraries in the comparisons as we did in Section \ref{sec:bench_perf} to make meaningful comparisons. Negative values for expressivity indicate negative outcomes: the gadget set in the debloated binary satisfies more expressivity classes than the original. The reverse is true for gadget set quality: positive values indicate that the average number of side constraints per gadget went down after debloating. However, it is worth noting that values less than 0.5 in either direction are not significant as GSA's quality scoring system values minor side constraints at 0.5 and major ones at 3.0. We averaged GSA's reported metrics across the three complexity levels, shown in Table \ref{tab:gadget_set_stats}. The first set of columns indicates the number of successfully debloated benchmarks for each tool at each complexity level.

% \usepackage{multirow}
% \usepackage{graphicx}
\begin{table*}[t]
\rowcolors{4}{lightgray!50}{white}
\centering
\caption{Average Changes in Gadget Set Expressivity and Quality }
\setlength{\tabcolsep}{6pt}
\label{tab:gadget_set_stats}
%\resizebox{.98\textwidth}{!}{%
\begin{threeparttable}
\begin{tabular}{|c|c|c|c||c|c|c||c|c|c|}
\cline{2-10}
\multicolumn{1}{c|}{} & \multicolumn{3}{c||}{\textbf{\# Benchmarks}} & \multicolumn{3}{c||}{\textbf{Gadget Set Expressivity}} & \multicolumn{3}{c|}{\textbf{Gadget Set Quality}} \\ \hline
\textbf{Debloater} & \textbf{Low} & \textbf{Medium} & \textbf{High} & \textbf{Low} & \textbf{Medium} & \textbf{High} & \textbf{Low} & \textbf{Medium} & \textbf{High} \\ \hline 
CHISEL & 9 & 6 & 2 & 1.3 & 0 & 0 & -0.1 & 0 & 0\\
CHISEL-GT & 8 & 3 & 2 & 0.3 & 0.7 & 0.5 & 0.1 & 0.2 & 0.1 \\
RAZOR & 10 & 3 & 2 & -0.3 & -0.3 & -0.5 & 0.2 & 0 & 0.1 \\
Binary Reduce (Dynamic) & 10 & 4 & 4 & 1.3 & 1 & 1.25 & 0.2 & 0.2 & 0.2\\ \hline
Trimmer (v2) [Agg.] & 7  & 4 & 4 & 0 & -0.3 & -3.5 & 0.4 & 0.2  & -0.3 \\
OCCAM (v2) [Agg.]  & 10 & 6 & 4 & 1.3 & 0.5 & 0.5 & 0.2 & 0 & 0.1\\
LMCAS-SIFT [Agg.]  & 7 & 2 & 1 & 0.1 & 0 & -1 & 0.2 & 0.2 & 0 \\ \hline
Binary Reduce (Static) & 10 & 6 & 1 & 0.3 & -0.5 & -1 & 0.3 & 0.1 & 0\\
Libfilter & 10 & 5 & 1 & 1 & 1 & 1 & 0 & 0 & 0 \\ \hline

\end{tabular}%
%\begin{tablenotes}
      %\scriptsize
%      \item * Indicates that at least one benchmark had negative results despite positive average change.
%    \end{tablenotes}
\end{threeparttable}
%}
\vspace{-.5em}
\end{table*}

Overall, the debloaters we evaluated did not have a significant impact on security posture according to these two metrics. In many cases, our data indicates expressivity and ease of chaining \emph{increased} for some tools and complexity levels. However, we consider only one average increase to be significant, which we define as a change of 2 or more expressivity classes or a change of 0.5 or more for the quality score. This case is TRIMMER's performance on the four high complexity benchmarks, where is increased expressivity by 3.5 classes on average. 

\textbf{Special Purpose Gadget Availability}. The third code re-usability metric we evaluated is concerned with the availability of special purpose code reuse gadgets in a program binary. Attackers use special purpose gadgets as scaffolding to organize functional gadgets or to accomplish specific tasks such as execute system calls (i.e., syscall gadget). The special purpose gadget availability metric captures how many different types of special purpose gadget types are available to an attacker, defined as the presence of at least one gadget per type. GSA defines 10 different types of special purpose gadgets, the most important of these types being syscall gadgets. Syscall gadgets are used to invoke operating system functions such as \texttt{exec} or \texttt{mprotect} on the attacker's behalf. These gadgets are rare outside of \texttt{libc}, as they only appear unintentionally in typical userland programs.

GSA compares the special purpose gadget types available in the debloated binary versus the original and reports the change in value. Negative values indicate negative outcomes: the gadget set in the debloated binary contains more types of special purpose gadgets than the original. Due to the relative importance of syscall gadgets, GSA also records the number of syscall gadgets present in the binary. We averaged GSA's reported metrics across the three complexity levels, shown in Table \ref{tab:sp_gadgets}. The first set of columns indicates the number of successfully debloated benchmarks for each tool at each complexity level. Additionally, the last set of columns displays the number of benchmarks where debloating introduced syscall gadgets where there were previously none (undesirable) or eliminated all available syscall gadgets (desirable). 

% \usepackage{multirow}
% \usepackage{graphicx}
\begin{table*}[t]
\rowcolors{4}{lightgray!50}{white}
\centering
\caption{Average Changes in Special Purpose Gadget Availability and Overall Syscall Gadget Impacts}
\setlength{\tabcolsep}{6pt}
\label{tab:sp_gadgets}
%\resizebox{.98\textwidth}{!}{%
\begin{threeparttable}
\begin{tabular}{|c|c|c|c||c|c|c||c|c|}
\cline{2-9}
\multicolumn{1}{c|}{} & \multicolumn{3}{c||}{\textbf{\# Benchmarks}} & \multicolumn{3}{c||}{\textbf{S.P. Gadget Types Available}} & \multicolumn{2}{c|}{\textbf{Syscall Gadget Impact}} \\ \hline
\textbf{Debloater} & \textbf{Low} & \textbf{Medium} & \textbf{High} & \textbf{Low} & \textbf{Medium} & \textbf{High} & \textbf{Eliminate} & \textbf{Introduce} \\ \hline 
CHISEL & 9 & 6 & 2 & 0.6 & -0.3 & 0 & 3 & 1 \\
CHISEL-GT & 8 & 3 & 2 & -0.6 & -2.3 & -1.5 & 1 & 3 \\
RAZOR & 10 & 3 & 2 & -0.9 & 0 & 0.5 & 1 & 2  \\
Binary Reduce (Dynamic) & 10 & 4 & 4 & -0.8 & -0.5 & 0.3 & 0 & 0 \\ \hline
Trimmer (v2) [Agg.] & 7  & 4 & 4 & 0.7 & 0.8 & -2.8 & 0 & 2   \\
OCCAM (v2) [Agg.]  & 10 & 6 & 4 & 0.4 & -0.5 & -0.3 & 1 & 0 \\
LMCAS-SIFT [Agg.]  & 7 & 2 & 1 & 4 & 3.5 & 2 & 0 & 0 \\ \hline
Binary Reduce (Static) & 10 & 6 & 1 & 2.9 & 2 & 1 & 0 & 0 \\
Libfilter & 10 & 5 & 1 & 0 & 0 & 0 & 0 & 0 \\ \hline

\end{tabular}%
%\begin{tablenotes}
      %\scriptsize
%      \item *  Indicates that at least one benchmark had negative results despite positive average change.
%    \end{tablenotes}
\end{threeparttable}
%}
\vspace{-.5em}
\end{table*}

Overall, debloating did not have a significant impact on the types of special purpose gadgets available. On average, the total number of types was increased or decreased by less than one type for almost all tools and complexity levels. CHISEL-GT and TRIMMER both introduced more than one type of special purpose gadget on average for medium+high and high complexity benchmarks, respectively. LMCAS-SIFT and Binary Reduce (Static) both reduce the availability of special purpose gadget types significantly at several levels of benchmark complexity, although it is important to note that this is due to the elimination of large amounts of code in libraries via static linking. Focusing specifically on syscall gadgets, the results were similarly mixed. Most tools either eliminated syscall gadgets for some benchmarks, introduced them in others, or more commonly, both.

\textbf{Gadget Locality}. The final code re-usability metric we evaluated is concerned with the portability of exploits written against the original program to debloated programs. Code re-use exploit payloads for methods such as return-oriented programming (ROP) and its variants reference gadgets selected by the attacker using their address in the binary. This exploit payload is fragile: it will only work on different versions of the same vulnerable program the attacker's gadgets can all be located at the prescribed addresses. While some highly complex exploitation techniques such as JIT-ROP \cite{jit-rop} can overcome the need for static gadget addresses, in practice intentionally randomizing the code layout in binaries provides for moving target defense and breaks portability of exploits. This forces the attacker to create an exploit payload for each different version of the vulnerable program they wish to exploit. GSA scans the original and debloated binaries to identify which gadgets retain their \emph{locality} after debloating, meaning the gadget remains at the same address in the binary and its instructions are not changed. GSA calculates locality as a percentage, where 0\% indicates there are no local gadgets in the debloated binary and 100\% indicates they are all local. 

Of the 136 successfully produced debloated variants, the majority of them (109) exhibited 0\% locality indicating they are immune to ported exploit payloads. These includes all of the debloated variants produced by LMCAS-SIFT, RAZOR, TRIMMER, and both SL debloaters. A further 18 debloated variants measured locality at less than 0.5\%, indicating strong protection from ported attacks. All debloated variants produced by OCCAM and CHISEL-GT are included in these two levels. The remaining 9 variants were produced by Binary Reduce (Dynamic) and CHISEL with gadget locality values in the range of 4\% to 82\%, indicating weak protection from ported exploits.

\subsection{Program Correctness / Robustness}

Debloating transformations are complex and ensuring they do not compromise the soundness of the program is challenging whether they are implemented pre- (i.e., S2S), during (i.e., CBS), or post-production (i.e., B2B, SL). Flawed or incomplete transformations have the potential to negatively impact programs in many ways that can manifest as logic bugs, run-time errors, crashes, and in some cases, introducing new vulnerabilities. We discovered ample evidence of this during our benchmark performance tests (Section \ref{sec:bench_perf}), where 36.6\% of the debloated binaries produced in our evaluation failed to execute their retained functionality.

This highlights the importance of post-debloating validation, a stage of the debloating workflow that is often neglected or left entirely to the user for tools targeting Type II bloat. This is partly due to a lack of effective testing tools; existing approaches like regression and fuzz testing do not naturally support testing debloated programs against their original versions. To address this shortcoming in our evaluation, we have created a differential testing tool for transformed programs called DIFFER that combines elements from differential, regression, and fuzz testing approaches.

DIFFER allows users to specify seed inputs that correspond to both retained and debloated features. It runs the original program and one or more of its debloated variants with these inputs and compares their outputs. DIFFER expects that inputs for retained features will result in outputs that are the same for the original and debloated programs. Conversely, it expects inputs for debloated features to cause the original and debloated programs to produce differing outputs. If DIFFER detects unexpected matches, differences or crashes it reports them to the user to inspect. DIFFER's reports can help the user identify mistakes in the debloater's configuration or instances where the debloated program is unsound. As is the case with all dynamic analysis tools, it is possible that DIFFER reports may be false positives. To reduce false positive rates to a minimum, DIFFER allows users to define custom output comparators that can account for expected differences in outputs (e.g., a program timestamps its console output). Additionally, DIFFER supports template-based mutational fuzzing of seed inputs to ensure maximum coverage of the input space (i.e., avoid false negatives) for both debloated and retained features.

It is important to note that DIFFER does not and cannot provide formal guarantees of soundness in debloaters or the debloated programs they produce. Like other dynamic analysis testing approaches, DIFFER cannot exhaustively test the input space for complex programs in the general case. Still, DIFFER is quite useful for post-debloating validation as it is user-friendly and only requires a moderate about of user expertise to configure. 

Using the general debloating specifications we created in Section \ref{sec:debloat-specs} as a starting point, we configured DIFFER to test both retained and debloated functions for each benchmark program. We ran DIFFER on each of the debloated benchmark programs that successfully completed their performance test (a total of 90 variants) for a maximum of 12 hours to identify crashes, inconsistencies, or errors introduced during debloating as well as failures to remove functionality marked for debloating. Our results are concerning: after filtering for false positives, DIFFER discovered errors or crashes in 27.8\% (25 of 90) of debloated benchmarks. Additionally, debloating tools failed to remove features marked for debloating in 60\% (39 of 65) of the remaining debloated benchmarks. Of the final 26 debloated benchmarks that passed testing by DIFFER, thirteen were produced by Binary Reduce (static), seven were produced by libfilter, four were produced by LMCAS-SIFT, and OCCAM and CHISEL each produced one. Across our evaluation, 50\% (20 of 40) of attempts to remove Type I bloat were ultimately successful, and \emph{only 3.3\%} of attempts to remove Type II bloat were successful (6 of 180)\footnote{Five successes were on low-complexity benchmarks, one was on a medium complexity benchmark.}. Our results indicate that the general lack of attention paid to post-debloating validation has resulted in over-reporting of successful debloating in the literature. This demonstrates the need for closer attention to post-debloating validation in future research to ensure accurate reporting of debloating success.

\section{Discussion and Key Findings}
\label{sec:discussion}

In this section we discuss our key findings and provide actionable recommendations for future debloating research.

\subsection{Low Tool Maturity}
Our most important finding is that debloating tools currently lack the maturity required to be used on real-world software. This is evidenced by a slim success rate of 42.5\% across all tools and complexity levels at producing debloated programs or libraries that can pass a performance test. The success rate across medium- and high-complexity benchmarks is only 22\%, indicating that the current generation of debloaters are suitable for use on lower complexity programs only. To advance the state of the art in debloating, new approaches should focus on supporting more complex software packages and programming language features / paradigms.

\subsection{Soundness Issues}
When tested for correctness and robustness with DIFFER, only two of the ten tools we evaluated, Binary Reduce static and dynamic, did not produce unsound debloated binaries (i.e., DIFFER did not detect any crashes, failures, or errors in kept functionality). In the case of Binary Reduce (dynamic), this tool produced only one debloated benchmark, and DIFFER did detect that it still contained features marked for debloating. As such, it is possible this binary is only sound because the tool did not successfully debloat any features, which is supported by the fact that this binary increased in size after debloating. 

Each of the remaining tools produced far more faulty programs than successful ones (except for BinRec-ToB, which did not produce any debloated binaries). While the majority of soundness issues detected by our performance tests and DIFFER were problems with retained features, our results also reveal serious program soundness issues with respect to debloated functionality. New approaches must evolve to cauterize the void left by excised code. In our survey, we observed only one tool with the capability to do so: CARVE \cite{carve}. CARVE introduces the concept of \emph{debloating with replacement} in which atomic (i.e., contiguous group of source code lines) debloating operations can be accompanied by replacement code that maintains the soundness and robustness of the program after debloating by inserting code to perform error handling, control-flow redirection, data-flow sanitization, etc. While promising, this feature requires manual inputs from developers and maintenance over time. A future research opportunity exists to explore automated and scalable methods for implementing debloating with replacement, potentially by re-purposing automated program repair techniques.

\subsection{Marginal Performance/Security Benefits}
Finally, the tools we evaluated demonstrated consistent and significant improvements in two metrics only: binary size and gadget locality. We observed marginal benefits at best for all other performance and security metrics. It is important to note that our security evaluation is limited to code re-usability metrics due to the difficulty of quantifying security improvements. These tools may have had positive impacts such as removing previously-unknown reachable vulnerabilities that are generally not possible to evaluate because they are undecidable.

Considering these marginal benefits in concert with the aforementioned maturity and soundness issues, users are likely to find the benefits of software debloating are currently outweighed by the costs and potential risks in real-world scenarios. Further research is needed to improve the benefits of debloating tools relative to the risks we identified above.

\section{Conclusion}
In this work, we presented our survey of the software debloating ecosystem. From this survey, we created a taxonomy of debloating tools, a collection of 12 metrics to measure their effectiveness, and a set of 20 benchmark programs of varying complexity useful for comparing tools against each other. Next, we evaluated and compared ten software debloating tools across the four most prominent types of debloating tools. Our evaluation measured the performance of the tools themselves, as well as the performance, soundness, and security improvements of the debloated programs they produce. Our evaluation indicates that the current generation of software debloating tools have several shortcomings that will prevent adoption of these techniques in real-world scenarios. Specifically, the software debloating tools we evaluated have limited support for medium- and high-complexity programs, struggle to maintain soundness and robustness during debloating, and have limited success in improving the performance or security posture of debloated programs. To drive adoption of software debloating tools, future debloating research should focus on improving support for complex programs, soundness, and metrics for quantifying positive security impacts. We have made our benchmark set, data, and custom tools publicly available to help drive debloater adoption and inform their future development.

%-------------------------------------------------------------------------------
\section{Availability}
\label{sec:availability}
%-------------------------------------------------------------------------------
We have made the full set of artifacts generated in this work
including our benchmark set, metric tools, DIFFER, 
evaluation scripts, and the evaluated tools publicly available at:

\begin{center}\url{https://github.com/trailofbits/debloater-eval}\end{center}
\begin{center}\url{https://github.com/trailofbits/differ}\end{center}

%-------------------------------------------------------------------------------
\section*{Acknowledgments}
%-------------------------------------------------------------------------------

This material is based upon work supported by the Office of Naval Research (ONR) under Contract No. N00014-21-C-1032.  Any opinions, findings and conclusions or recommendations expressed in this material are those of the author(s) and do not necessarily reflect the views of the ONR. The authors would also like to thank Nathan Harris for their assistance in evaluating tools for this study.

%\begin{comment}
%\appendix
%\section*{Appendices}

%\end{comment}

%-------------------------------------------------------------------------------
\bibliographystyle{plain}
\bibliography{refs}

\begin{thebibliography}{10}

\bibitem{nibbler}
Ioannis Agadakos, Nicholas Demarinis, Di~Jin, Kent Williams-King, Jearson
  Alfajardo, Benjamin Shteinfeld, David Williams-King, Vasileios~P Kemerlis,
  and Georgios Portokalidis.
\newblock Large-scale debloating of binary shared libraries.
\newblock {\em Digital Threats: Research and Practice}, 1(4):1--28, 2020.

\bibitem{trimmer2}
Aatira~Anum Ahmad, Abdul~Rafae Noor, Hashim Sharif, Usama Hameed, Shoaib Asif,
  Mubashir Anwar, Ashish Gehani, Fareed Zaffar, and Junaid~Haroon Siddiqui.
\newblock Trimmer: an automated system for configuration-based software
  debloating.
\newblock {\em IEEE Transactions on Software Engineering}, 48(9):3485--3505,
  2021.

\bibitem{lmcas}
Mohannad Alhanahnah, Rithik Jain, Vaibhav Rastogi, Somesh Jha, and Thomas Reps.
\newblock Lightweight, multi-stage, compiler-assisted application
  specialization.
\newblock In {\em 2022 IEEE 7th European Symposium on Security and Privacy
  (EuroS\&P)}, pages 251--269. IEEE, 2022.

\bibitem{debloatbencha}
Muaz Ali, Muhammad Muzammil, Faraz Karim, Ayesha Naeem, Rukhshan Haroon,
  Muhammad Haris, Huzaifah Nadeem, Waseem Sabir, Fahad Shaon, Fareed Zaffar,
  et~al.
\newblock Sok: A tale of reduction, security, and correctness-evaluating
  program debloating paradigms and their compositions.
\newblock ESORICS, 2023.

\bibitem{binrec}
Anil Altinay, Joseph Nash, Taddeus Kroes, Prabhu Rajasekaran, Dixin Zhou,
  Adrian Dabrowski, David Gens, Yeoul Na, Stijn Volckaert, Cristiano Giuffrida,
  et~al.
\newblock Binrec: dynamic binary lifting and recompilation.
\newblock In {\em Proceedings of the Fifteenth European Conference on Computer
  Systems}, pages 1--16, 2020.

\bibitem{chiselbench}
aspire project.
\newblock Chiselbench.
\newblock \url{https://github.com/aspire-project/chisel-bench}, 2023.

\bibitem{animatedead}
Babak~Amin Azad, Rasoul Jahanshahi, Chris Tsoukaladelis, Manuel Egele, and Nick
  Nikiforakis.
\newblock $\{$AnimateDead$\}$: Debloating web applications using concolic
  execution.
\newblock In {\em 32nd USENIX Security Symposium (USENIX Security 23)}, pages
  5575--5591, 2023.

\bibitem{lessismore}
Babak~Amin Azad, Pierre Laperdrix, and Nick Nikiforakis.
\newblock Less is more: quantifying the security benefits of debloating web
  applications.
\newblock In {\em 28th USENIX Security Symposium (USENIX Security 19)}, pages
  1697--1714, 2019.

\bibitem{ancile}
Priyam Biswas, Nathan Burow, and Mathias Payer.
\newblock Code specialization through dynamic feature observation.
\newblock In {\em Proceedings of the Eleventh ACM Conference on Data and
  Application Security and Privacy}, pages 257--268, 2021.

\bibitem{dacapo}
Stephen~M Blackburn, Robin Garner, Chris Hoffmann, Asjad~M Khang, Kathryn~S
  McKinley, Rotem Bentzur, Amer Diwan, Daniel Feinberg, Daniel Frampton,
  Samuel~Z Guyer, et~al.
\newblock The dacapo benchmarks: Java benchmarking development and analysis.
\newblock In {\em Proceedings of the 21st annual ACM SIGPLAN conference on
  Object-oriented programming systems, languages, and applications}, pages
  169--190, 2006.

\bibitem{carve}
Michael~D Brown and Santosh Pande.
\newblock Carve: Practical security-focused software debloating using simple
  feature set mappings.
\newblock In {\em Proceedings of the 3rd ACM Workshop on Forming an Ecosystem
  Around Software Transformation}, pages 1--7, 2019.

\bibitem{GSA}
Michael~D Brown and Santosh Pande.
\newblock Is less really more? towards better metrics for measuring security
  improvements realized through software debloating.
\newblock In {\em 12th USENIX Workshop on Cyber Security Experimentation and
  Test (CSET 19)}, 2019.

\bibitem{jshrink}
Bobby~R Bruce, Tianyi Zhang, Jaspreet Arora, Guoqing~Harry Xu, and Miryung Kim.
\newblock Jshrink: In-depth investigation into debloating modern java
  applications.
\newblock In {\em Proceedings of the 28th ACM joint meeting on european
  software engineering conference and symposium on the foundations of software
  engineering}, pages 135--146, 2020.

\bibitem{deadmethods}
Danilo Caivano, Pietro Cassieri, Simone Romano, and Giuseppe Scanniello.
\newblock An exploratory study on dead methods in open-source java desktop
  applications.
\newblock In {\em Proceedings of the 15th ACM/IEEE International Symposium on
  Empirical Software Engineering and Measurement (ESEM)}, pages 1--11, 2021.

\bibitem{toss}
Yurong Chen, Shaowen Sun, Tian Lan, and Guru Venkataramani.
\newblock Toss: Tailoring online server systems through binary feature
  customization.
\newblock In {\em Proceedings of the 2018 Workshop on Forming an Ecosystem
  Around Software Transformation}, pages 1--7, 2018.

\bibitem{decaf}
Jake Christensen, Ionut~Mugurel Anghel, Rob Taglang, Mihai Chiroiu, and Radu
  Sion.
\newblock $\{$DECAF$\}$: Automatic, adaptive de-bloating and hardening of
  $\{$COTS$\}$ firmware.
\newblock In {\em 29th USENIX Security Symposium (USENIX Security 20)}, pages
  1713--1730, 2020.

\bibitem{mitre-cve}
MITRE Corporation.
\newblock Cve database.
\newblock \url{https://cve.mitre.org/}, 2023.

\bibitem{spec2006}
Standard Performance~Evaluation Corporation.
\newblock Spec cpu 2006.
\newblock \url{https://www.spec.org/cpu2006/}, 2018.

\bibitem{spec2017}
Standard Performance~Evaluation Corporation.
\newblock Spec cpu 2017.
\newblock \url{https://www.spec.org/cpu2017/}, 2023.

\bibitem{cbat}
Michael Crystal, Chris Casinghino, and Inc. The Charles Stark
  Draper~Laboratory.
\newblock Cbat: A comparative binary analysis tool.
\newblock 2022.

\bibitem{sludge}
Josiah Dykstra, Kelly Shortridge, Jamie Met, and Douglas Hough.
\newblock Sludge for good: Slowing and imposing costs on cyber attackers.
\newblock {\em arXiv preprint arXiv:2211.16626}, 2022.

\bibitem{binarycontrolflowtrimming}
Masoud Ghaffarinia and Kevin~W Hamlen.
\newblock Binary control-flow trimming.
\newblock In {\em Proceedings of the 2019 ACM SIGSAC Conference on Computer and
  Communications Security}, pages 1009--1022, 2019.

\bibitem{confine}
Seyedhamed Ghavamnia, Tapti Palit, Azzedine Benameur, and Michalis
  Polychronakis.
\newblock Confine: Automated system call policy generation for container attack
  surface reduction.
\newblock In {\em 23rd International Symposium on Research in Attacks,
  Intrusions and Defenses (RAID 2020)}, pages 443--458, 2020.

\bibitem{C2C}
Seyedhamed Ghavamnia, Tapti Palit, and Michalis Polychronakis.
\newblock C2c: Fine-grained configuration-driven system call filtering.
\newblock In {\em Proceedings of the 2022 ACM SIGSAC Conference on Computer and
  Communications Security}, pages 1243--1257, 2022.

\bibitem{coreutils}
GNU.
\newblock Gnu coreutils.
\newblock \url{https://www.gnu.org/software/coreutils/}, 2023.

\bibitem{binary-reduce}
Inc. Grammatech.
\newblock Binary reduce.
\newblock \url{https://grammatech.github.io/prj/binary-reduce/}, 2022.

\bibitem{facechange}
Zhongshu Gu, Brendan Saltaformaggio, Xiangyu Zhang, and Dongyan Xu.
\newblock Face-change: Application-driven dynamic kernel view switching in a
  virtual machine.
\newblock In {\em 2014 44th Annual IEEE/IFIP International Conference on
  Dependable Systems and Networks}, pages 491--502. IEEE, 2014.

\bibitem{proguard}
Guardsquare.
\newblock Proguard.
\newblock \url{https://github.com/Guardsquare/proguard}, 2023.

\bibitem{debloatbenchc}
Muhammad Hassan, Talha Tahir, Muhammad Farrukh, Abdullah Naveed, Anas Naeem,
  Fahad Shaon, Fareed Zaffar, Ashish Gehani, and Sazzadur Rahaman.
\newblock Evaluating container debloaters.
\newblock In {\em IEEE Secure Development Conference, SecDev}, pages 18--20,
  2023.

\bibitem{polydroid}
Brian Heath, Neelay Velingker, Osbert Bastani, and Mayur Naik.
\newblock Polydroid: Learning-driven specialization of mobile applications.
\newblock {\em arXiv preprint arXiv:1902.09589}, 2019.

\bibitem{chisel}
Kihong Heo, Woosuk Lee, Pardis Pashakhanloo, and Mayur Naik.
\newblock Effective program debloating via reinforcement learning.
\newblock In {\em Proceedings of the 2018 ACM SIGSAC Conference on Computer and
  Communications Security}, pages 380--394, 2018.

\bibitem{minimalist}
Rasoul Jahanshahi, Babak~Amin Azad, Nick Nikiforakis, and Manuel Egele.
\newblock Minimalist: Semi-automated debloating of $\{$PHP$\}$ web applications
  through static analysis.
\newblock In {\em 32nd USENIX Security Symposium (USENIX Security 23)}, pages
  5557--5573, 2023.

\bibitem{reddroid}
Yufei Jiang, Qinkun Bao, Shuai Wang, Xiao Liu, and Dinghao Wu.
\newblock Reddroid: Android application redundancy customization based on
  static analysis.
\newblock In {\em 2018 IEEE 29th international symposium on software
  reliability engineering (ISSRE)}, pages 189--199. IEEE, 2018.

\bibitem{jred}
Yufei Jiang, Dinghao Wu, and Peng Liu.
\newblock Jred: Program customization and bloatware mitigation based on static
  analysis.
\newblock In {\em 2016 IEEE 40th annual computer software and applications
  conference (COMPSAC)}, volume~1, pages 12--21. IEEE, 2016.

\bibitem{jreduce}
Christian~Gram Kalhauge and Jens Palsberg.
\newblock Binary reduction of dependency graphs.
\newblock In {\em Proceedings of the 2019 27th ACM Joint Meeting on European
  Software Engineering Conference and Symposium on the Foundations of Software
  Engineering}, pages 556--566, 2019.

\bibitem{mininode}
Igibek Koishybayev and Alexandros Kapravelos.
\newblock Mininode: Reducing the attack surface of node. js applications.
\newblock In {\em 23rd International Symposium on Research in Attacks,
  Intrusions and Defenses (RAID 2020)}, pages 121--134, 2020.

\bibitem{config-driven}
Hyungjoon Koo, Seyedhamed Ghavamnia, and Michalis Polychronakis.
\newblock Configuration-driven software debloating.
\newblock In {\em Proceedings of the 12th European Workshop on Systems
  Security}, pages 1--6, 2019.

\bibitem{kulkarni2009practical}
Prasad~A Kulkarni, David~B Whalley, Gary~S Tyson, and Jack~W Davidson.
\newblock Practical exhaustive optimization phase order exploration and
  evaluation.
\newblock {\em ACM Transactions on Architecture and Code Optimization (TACO)},
  6(1):1--36, 2009.

\bibitem{kulkarni2012mitigating}
Sameer Kulkarni and John Cavazos.
\newblock Mitigating the compiler optimization phase-ordering problem using
  machine learning.
\newblock In {\em Proceedings of the ACM international conference on Object
  oriented programming systems languages and applications}, pages 147--162,
  2012.

\bibitem{cozart}
Hsuan-Chi Kuo, Jianyan Chen, Sibin Mohan, and Tianyin Xu.
\newblock Set the configuration for the heart of the os: On the practicality of
  operating system kernel debloating.
\newblock {\em Proceedings of the ACM on Measurement and Analysis of Computing
  Systems}, 4(1):1--27, 2020.

\bibitem{deepoccam}
Nham Le~Van, Ashish Gehani, Arie Gurfinkel, Susmit Jha, and Jorge~A Navas.
\newblock Reinforcement learning guided software debloating.

\bibitem{occam}
Gregory Malecha, Ashish Gehani, and Natarajan Shankar.
\newblock Automated software winnowing.
\newblock In {\em Proceedings of the 30th Annual ACM Symposium on Applied
  Computing}, pages 1504--1511, 2015.

\bibitem{fblocker}
Mohamad Mansouri, Jun Xu, and Georgios Portokalidis.
\newblock Eliminating vulnerabilities by disabling unwanted functionality in
  binary programs.
\newblock In {\em Proceedings of the 2023 ACM Asia Conference on Computer and
  Communications Security}, pages 259--273, 2023.

\bibitem{leansym}
Xianya Mi, Sanjay Rawat, Cristiano Giuffrida, and Herbert Bos.
\newblock Leansym: Efficient hybrid fuzzing through conservative constraint
  debloating.
\newblock In {\em Proceedings of the 24th International Symposium on Research
  in Attacks, Intrusions and Defenses}, pages 62--77, 2021.

\bibitem{occam2}
Jorge~A Navas and Ashish Gehani.
\newblock Occam-v2: Combining static and dynamic analysis for effective and
  efficient whole-program specialization.
\newblock {\em Communications of the ACM}, 66(4):40--47, 2023.

\bibitem{cachetor}
Khanh Nguyen and Guoqing Xu.
\newblock Cachetor: Detecting cacheable data to remove bloat.
\newblock In {\em Proceedings of the 2013 9th Joint Meeting on Foundations of
  Software Engineering}, pages 268--278, 2013.

\bibitem{pacjam}
Pardis Pashakhanloo, Aravind Machiry, Hyonyoung Choi, Anthony Canino, Kihong
  Heo, Insup Lee, and Mayur Naik.
\newblock Pacjam: Securing dependencies continuously via package-oriented
  debloating.
\newblock In {\em Proceedings of the 2022 ACM on Asia Conference on Computer
  and Communications Security}, pages 903--916, 2022.

\bibitem{depclean}
Serena~Elisa Ponta, Wolfram Fischer, Henrik Plate, and Antonino Sabetta.
\newblock The used, the bloated, and the vulnerable: Reducing the attack
  surface of an industrial application.
\newblock In {\em 2021 IEEE International Conference on Software Maintenance
  and Evolution (ICSME)}, pages 555--558. IEEE, 2021.

\bibitem{decker}
Chris Porter, Sharjeel Khan, and Santosh Pande.
\newblock Decker: Attack surface reduction via on-demand code mapping.
\newblock In {\em Proceedings of the 28th ACM International Conference on
  Architectural Support for Programming Languages and Operating Systems, Volume
  2}, pages 192--206, 2023.

\bibitem{blankit}
Chris Porter, Girish Mururu, Prithayan Barua, and Santosh Pande.
\newblock Blankit library debloating: Getting what you want instead of cutting
  what you don’t.
\newblock In {\em Proceedings of the 41st ACM SIGPLAN Conference on Programming
  Language Design and Implementation}, pages 164--180, 2020.

\bibitem{razor}
Chenxiong Qian, Hong Hu, Mansour Alharthi, Pak~Ho Chung, Taesoo Kim, and Wenke
  Lee.
\newblock $\{$RAZOR$\}$: A framework for post-deployment software debloating.
\newblock In {\em 28th USENIX security symposium (USENIX Security 19)}, pages
  1733--1750, 2019.

\bibitem{slimium}
Chenxiong Qian, Hyungjoon Koo, ChangSeok Oh, Taesoo Kim, and Wenke Lee.
\newblock Slimium: debloating the chromium browser with feature subsetting.
\newblock In {\em Proceedings of the 2020 ACM SIGSAC Conference on Computer and
  Communications Security}, pages 461--476, 2020.

\bibitem{multi-os}
Anh Quach, Rukayat Erinfolami, David Demicco, and Aravind Prakash.
\newblock A multi-os cross-layer study of bloating in user programs, kernel and
  managed execution environments.
\newblock In {\em Proceedings of the 2017 Workshop on Forming an Ecosystem
  Around Software Transformation}, pages 65--70, 2017.

\bibitem{bloatfactors}
Anh Quach and Aravind Prakash.
\newblock Bloat factors and binary specialization.
\newblock In {\em Proceedings of the 3rd ACM Workshop on Forming an Ecosystem
  Around Software Transformation}, pages 31--38, 2019.

\bibitem{pcl}
Anh Quach, Aravind Prakash, and Lok Yan.
\newblock Debloating software through $\{$Piece-Wise$\}$ compilation and
  loading.
\newblock In {\em 27th USENIX security symposium (USENIX Security 18)}, pages
  869--886, 2018.

\bibitem{cimplifier}
Vaibhav Rastogi, Drew Davidson, Lorenzo De~Carli, Somesh Jha, and Patrick
  McDaniel.
\newblock Cimplifier: automatically debloating containers.
\newblock In {\em Proceedings of the 2017 11th Joint Meeting on Foundations of
  Software Engineering}, pages 476--486, 2017.

\bibitem{BinTrimmer}
Nilo Redini, Ruoyu Wang, Aravind Machiry, Yan Shoshitaishvili, Giovanni Vigna,
  and Christopher Kruegel.
\newblock B in t rimmer: Towards static binary debloating through abstract
  interpretation.
\newblock In {\em Detection of Intrusions and Malware, and Vulnerability
  Assessment: 16th International Conference, DIMVA 2019, Gothenburg, Sweden,
  June 19--20, 2019, Proceedings 16}, pages 482--501. Springer, 2019.

\bibitem{creduce}
John Regehr, Yang Chen, Pascal Cuoq, Eric Eide, Chucky Ellison, and Xuejun
  Yang.
\newblock Test-case reduction for c compiler bugs.
\newblock In {\em Proceedings of the 33rd ACM SIGPLAN conference on Programming
  Language Design and Implementation}, pages 335--346, 2012.

\bibitem{rewriter-eval}
Eric Schulte, Michael~D Brown, and Vlad Folts.
\newblock A broad comparative evaluation of x86-64 binary rewriters.
\newblock In {\em Proceedings of the 15th Workshop on Cyber Security
  Experimentation and Test}, pages 129--144, 2022.

\bibitem{gtirb}
Eric Schulte, Jonathan Dorn, Antonio Flores-Montoya, Aaron Ballman, and Tom
  Johnson.
\newblock Gtirb: intermediate representation for binaries.
\newblock {\em arXiv preprint arXiv:1907.02859}, 2019.

\bibitem{trimmer}
Hashim Sharif, Muhammad Abubakar, Ashish Gehani, and Fareed Zaffar.
\newblock Trimmer: application specialization for code debloating.
\newblock In {\em Proceedings of the 33rd ACM/IEEE International Conference on
  Automated Software Engineering}, pages 329--339, 2018.

\bibitem{libfilter}
Benjamin Shteinfeld.
\newblock Libfilter: Debloating dynamically-linked libraries through binary
  recompilation.
\newblock {\em Undergraduate Honors Thesis. Brown University}, 2019.

\bibitem{jit-rop}
Kevin~Z Snow, Fabian Monrose, Lucas Davi, Alexandra Dmitrienko, Christopher
  Liebchen, and Ahmad-Reza Sadeghi.
\newblock Just-in-time code reuse: On the effectiveness of fine-grained address
  space layout randomization.
\newblock In {\em 2013 IEEE symposium on security and privacy}, pages 574--588.
  IEEE, 2013.

\bibitem{finegrainlibrary}
Linhai Song and Xinyu Xing.
\newblock Fine-grained library customization.
\newblock {\em arXiv preprint arXiv:1810.11128}, 2018.

\bibitem{cov-based}
C{\'e}sar Soto-Valero, Thomas Durieux, Nicolas Harrand, and Benoit Baudry.
\newblock Coverage-based debloating for java bytecode.
\newblock {\em ACM Transactions on Software Engineering and Methodology},
  32(2):1--34, 2023.

\bibitem{occambench}
SRI-CSL.
\newblock Occam-benchmarks.
\newblock
  \url{https://github.com/SRI-CSL/OCCAM-Benchmarks/tree/master/examples}, 2023.

\bibitem{perses}
Chengnian Sun, Yuanbo Li, Qirun Zhang, Tianxiao Gu, and Zhendong Su.
\newblock Perses: Syntax-guided program reduction.
\newblock In {\em Proceedings of the 40th International Conference on Software
  Engineering}, pages 361--371, 2018.

\bibitem{xdebloat}
Yutian Tang, Hao Zhou, Xiapu Luo, Ting Chen, Haoyu Wang, Zhou Xu, and Yan Cai.
\newblock Xdebloat: Towards automated feature-oriented app debloating.
\newblock {\em IEEE Transactions on Software Engineering}, 48(11):4501--4520,
  2021.

\bibitem{jax}
Frank Tip, Chris Laffra, Peter~F Sweeney, and David Streeter.
\newblock Practical experience with an application extractor for java.
\newblock In {\em Proceedings of the 14th ACM SIGPLAN conference on
  Object-oriented programming, systems, languages, and applications}, pages
  292--305, 1999.

\bibitem{binrec-repo}
Inc. Trail~of Bits.
\newblock Binrec-tob.
\newblock \url{https://github.com/trailofbits/binrec-tob}, 2022.

\bibitem{stubbifier}
Alexi Turcotte, Ellen Arteca, Ashish Mishra, Saba Alimadadi, and Frank Tip.
\newblock Stubbifier: debloating dynamic server-side javascript applications.
\newblock {\em Empirical Software Engineering}, 27(7):161, 2022.

\bibitem{prat}
Ryan Williams, Tongwei Ren, Lorenzo De~Carli, Long Lu, and Gillian Smith.
\newblock Guided feature identification and removal for resource-constrained
  firmware.
\newblock {\em ACM Transactions on Software Engineering and Methodology
  (TOSEM)}, 31(2):1--25, 2021.

\bibitem{egalito}
David Williams-King, Hidenori Kobayashi, Kent Williams-King, Graham Patterson,
  Frank Spano, Yu~Jian Wu, Junfeng Yang, and Vasileios~P Kemerlis.
\newblock Egalito: Layout-agnostic binary recompilation.
\newblock In {\em Proceedings of the Twenty-Fifth International Conference on
  Architectural Support for Programming Languages and Operating Systems}, pages
  133--147, 2020.

\bibitem{debop}
Qi~Xin, Myeongsoo Kim, Qirun Zhang, and Alessandro Orso.
\newblock Program debloating via stochastic optimization.
\newblock In {\em Proceedings of the ACM/IEEE 42nd International Conference on
  Software Engineering: New Ideas and Emerging Results}, pages 65--68, 2020.

\bibitem{domgad}
Qi~Xin, Myeongsoo Kim, Qirun Zhang, and Alessandro Orso.
\newblock Subdomain-based generality-aware debloating.
\newblock In {\em Proceedings of the 35th IEEE/ACM International Conference on
  Automated Software Engineering}, pages 224--236, 2020.

\bibitem{covaf}
Qi~Xin, Qirun Zhang, and Alessandro Orso.
\newblock Studying and understanding the tradeoffs between generality and
  reduction in software debloating.
\newblock In {\em Proceedings of the 37th IEEE/ACM International Conference on
  Automated Software Engineering}, pages 1--13, 2022.

\bibitem{hecate}
Hongfa Xue, Yurong Chen, Guru Venkataramani, and Tian Lan.
\newblock Hecate: Automated customization of program and communication features
  to reduce attack surfaces.
\newblock In {\em Security and Privacy in Communication Networks: 15th EAI
  International Conference, SecureComm 2019, Orlando, FL, USA, October 23--25,
  2019, Proceedings, Part II 15}, pages 305--319. Springer, 2019.

\bibitem{jslim}
Renjun Ye, Liang Liu, Simin Hu, Fangzhou Zhu, Jingxiu Yang, and Feng Wang.
\newblock Jslim: Reducing the known vulnerabilities of javascript application
  by debloating.
\newblock In {\em International Symposium on Emerging Information Security and
  Applications}, pages 128--143. Springer, 2021.

\bibitem{utrimmer}
Haotian Zhang, Mengfei Ren, Yu~Lei, and Jiang Ming.
\newblock One size does not fit all: security hardening of mips embedded
  systems via static binary debloating for shared libraries.
\newblock In {\em Proceedings of the 27th ACM International Conference on
  Architectural Support for Programming Languages and Operating Systems}, pages
  255--270, 2022.

\bibitem{asandebloat}
Yuchen Zhang, Chengbin Pang, Georgios Portokalidis, Nikos Triandopoulos, and
  Jun Xu.
\newblock Debloating address sanitizer.
\newblock In {\em 31st USENIX Security Symposium (USENIX Security 22)}, pages
  4345--4363, 2022.

\bibitem{lightweightbinarytailoring}
Andreas Ziegler, Julian Geus, Bernhard Heinloth, Timo H{\"o}nig, and Daniel
  Lohmann.
\newblock Honey, i shrunk the elfs: Lightweight binary tailoring of shared
  libraries.
\newblock {\em ACM Transactions on Embedded Computing Systems (TECS)},
  18(5s):1--23, 2019.

\end{thebibliography}

%%%%%%%%%%%%%%%%%%%%%%%%%%%%%%%%%%%%%%%%%%%%%%%%%%%%%%%%%%%%%%%%%%%%%%%%%%%%%%%%
\end{document}